\documentclass{iau}

\usepackage{amsmath}
\usepackage{graphicx}
\usepackage{multirow}

\begin{document}

\lefttitle{Yuzhe Song et. al}
\righttitle{Modelling Millisecond Pulsar Populations in Globular Clusters with NBODY6++GPU}

\jnlPage{1}{7}
\jnlDoiYr{2021}
\doival{10.1017/xxxxx}

\aopheadtitle{Proceedings IAU Symposium}
\editors{Editors.}

\title{Modelling Millisecond Pulsar Populations in Globular Clusters with NBODY6++GPU}

\author{Yuzhe Song$^{1,2}$, Debatri Chattopadhyay$^{3}$, Jarrod Hurley$^{1,2}$, \\
Rainer Spurzem$^{4,5,6}$, Francesco Flammini Dotti$^{7,4}$ and Kai Wu$^{4}$}
\affiliation{$^{1}$Swinburne University of Technology, $^{2}$OzGrav, $^{3}$Northwestern University, $^{4}$Heidelberg University, $^{5}$Peking University, $^{6}$National Astronomical Observatories, China, $^{7}$New York University Abu Dhabi}

\begin{abstract}

Millisecond pulsars (MSPs) are neutron stars with spin periods as short as a few milliseconds, formed through mass accretion from companion stars.
In the dense environments of globular clusters (GCs), MSPs are likely to originate through dynamically assembled interacting binaries.
Over 300 MSPs have been detected in GCs to date, more than half of the known Galactic MSP population.
In this work, we model MSP populations in intermediate-mass GCs using the direct $N$-body code \textsc{NBODY6++GPU}.
We update the code by implementing pulsar spin-down due to magnetic braking and spin-up through accretion, and use this framework to model the pulsar population in the globular cluster M71 to investigate the pulsar population within and the associated gravitational wave transients.

\end{abstract}


\maketitle

\section{Introduction}

Pulsars are rapidly rotating neutron stars associated with a wide range of high-energy astrophysical phenomena.  
Those in binary systems are of particular interest: they can be spun up into millisecond pulsars (MSPs), form double compact objects that eventually merge, and contribute to the dynamical evolution of their host systems.  
Despite their importance, most direct $N$-body codes do not explicitly model pulsar evolution.  
The only exception to date is \citet{Ye2019ApJ}, which adopted a highly simplified prescription.  
Here, we implement a more realistic treatment of pulsar spin and magnetic field evolution \citep{Debatri2020, Song2024} into the state-of-the-art direct $N$-body code \textsc{NBODY6++GPU} \citep{Wang2015, Kamlah2022, Spurzem2023}.  

Globular clusters (GCs), containing $\sim 10^6$ stars in dense stellar environments, are natural laboratories for studying pulsar formation and evolution.  
Observations with \textit{Fermi}-LAT have detected $\gamma$-ray emission from 36 of the 154 known Galactic GCs, and recent stacking analyses \citep[e.g.,][]{Henry2024} suggest that even more clusters may emit at these energies.  
The $\gamma$-ray emission is thought to originate primarily from MSPs, either through the superposition of unresolved pulsars or via inverse Compton scattering between relativistic pulsar winds and ambient soft photons.  

The number and properties of MSPs in clusters are expected to depend on the stellar encounter rate \citep{Tauris2006}, linking their $\gamma$-ray output directly to cluster dynamics.  
In this work, we use direct $N$-body simulations to investigate how GC dynamical evolution shapes the MSP population and, consequently, the $\gamma$-ray emission of clusters.  

\section{Pulsar Evolution Model}

In this section we introduce the pulsar evolution models implemented in \textsc{NBODY6++GPU}. 

\subsection{Evolution of Isolated Pulsars}
Isolated pulsars are assumed to spin down via magnetic braking with braking index $n=3$.
Their surface magnetic fields may decay over time, though the decay timescale remains uncertain.
Field decay is thought to arise from the Hall effect, which redistributes magnetic energy and enhances Ohmic dissipation \citep{Romani:1990Natur, Konar:1997MNRAS.284..311K}.  
In this work we adopt a simple exponential decay model, consistent with previous studies \citep{Kiel:2008MNRAS,Debatri2020,Dirson2022,Song2024}: 
\begin{equation}
    B_f = (B_i - B_{\textrm{min}}) \exp(-t / \tau) + B_{\textrm{min}},
	\label{eq:isolate_B}
\end{equation}
where $B_i$ and $B_f$ are the initial and final magnetic fields, $B_{\textrm{min}} = 10^{8}$\,G \citep{Zhang:2006MNRAS}, and $\tau$ is the decay timescale.  

The spin period derivative, assuming vacuum dipole radiation, is

\begin{equation}
    \dot{P} = \frac{8 \pi ^ 2  R ^ 6 B^2 \sin^2{\alpha}}{3 c^3 I P} ,
    \label{eq:isolate_Pdot}
\end{equation}
where $R=10$\,km, $c$ is the speed of light, $\alpha$ is the inclination angle, and the NS moment of inertia is \citep{lattimer2005}
\begin{equation}
    I = (0.237 \pm 0.008) M R^2 [1 + 4.2\frac{\textrm{M}}{\textrm{M}_{\odot}} \frac{\textrm{km}}{\textrm{R}} + 90(\frac{\textrm{M}}{\textrm{M}_{\odot}}\frac{\textrm{km}}{\textrm{R}})^4] ,
    \label{eq:moment_of_inertia}
\end{equation}
Integrating Eq.~\ref{eq:isolate_Pdot} with $B(t)$ from Eq.~\ref{eq:isolate_B} gives
\begin{equation}
    P^2 = \frac{16 \pi ^ 2  R ^ 6 \sin^2{\alpha}}{3 c^3 I}  [B_{\textrm{min}}^2 t - \tau B_{\textrm{min}} (B_f - B_i) - \frac{\tau}{2}(B^2_f - B^2_i)] + P_0^2 .
    \label{eq:isolate_P}
\end{equation}
where $P_0$ is the initial spin period.  
The evolution of $P$ depends on $\alpha$. 
Although $\alpha$ may evolve with $B$ \citep{Dirson2022}, here we assume a fixed $\alpha$, decoupling its evolution from $P$ and $B$.  
Observations suggest older pulsars align their axes \citep{Young:2010MNRAS}, but modelling of radio and \gray\ pulsars favours large $\alpha$ values \citep{Johnston2020}.  
Accordingly, we set $\alpha = \pi/2$. 

\subsection{Evolution of Pulsars in Roche Lobe Overflow}

Roche lobe overflow (RLOF) occurs when a star fills its Roche lobe and transfers mass through the inner Lagrangian point, often forming an accretion disk. The Roche lobe radius is approximated by \citet{1983ApJ...268..368E}:

\begin{equation}
   \frac{ R_{\rm L}}{a} = \frac{0.49q^{2/3}}{0.6q^{2/3} + \ln{(1+q^{1/3})}},
   \label{eq:eggleton_RL}
\end{equation}
where $R_{\rm L}$ is the Roche lobe radius, $a$ the orbital separation, and $q = M_{\rm comp}/M_{\rm NS}$ the companion-to-NS mass ratio.  

Accretion during RLOF drives (i) magnetic field decay through burial and (ii) NS spin-up. Field decay follows

\begin{equation}
    B = (B_0 - B_{\textrm{min}}) \exp(-\Delta M / \Delta M_d) + B_{\textrm{min}},
	\label{eq:recycle_B}
\end{equation}

where $\Delta M$ is the accreted mass, $\Delta M_d$ a characteristic mass scale, $B_0$ the initial field, and $B_{\rm min}$ the minimum allowed field. The rate of angular momentum increase during accretion is 

\begin{equation}
    \dot{J}_\textbf{\textrm{acc}} = \epsilon V_{\textrm{diff}} R^2_A \dot{M}_{\textrm{NS}}, 
    \label{eq:jdot}
\end{equation}
where $\epsilon=1.0$ is the efficiency factor, $\dot{M}_{\textrm{NS}}$ the accretion rate, $V_{\textrm{diff}}$ the difference between the Keplerian angular velocity at the magnetic radius and the co-rotation velocity, and $R_A$ half the Alfvén radius:

\begin{equation}
    R_{\rm Alfven} = (\frac{8 R^{12} B^4}{M \dot{M}^2 G})^{\frac{1}{7}}.
\end{equation}

The final spin period after mass transfer is

\begin{equation}
    P_{\rm f} = 2 \pi / (2 \pi/P_{\rm i} + \Delta J_{\rm acc} / I ),
\end{equation}
where $P_{\rm i}$ is the initial spin period, $\Delta J_{\rm acc}$ the angular momentum gained, and $I$ the moment of inertia (Eq.~\ref{eq:moment_of_inertia}).  
The corresponding $\dot{P}$ is computed following \citet{Song2024}.

\subsection{Evolution of Pulsars in Common Envelope}
\label{subsec:ce}

Common envelope (CE) evolution occurs when the donor engulfs its companion in a shared envelope, typically leading to dramatic orbital shrinkage. CE evolution is critical for forming tight binaries that may merge as gravitational-wave sources. NS behaviour during CE is uncertain, so we implement three scenarios. 

(1) No accretion during CE; NS spins down naturally.

(2) NS accretes via an accretion disk, as in RLOF, allowing spin-up.
  
(3) Direct surface accretion onto the NS with spin-up, where $V_{\textrm{diff}}$ in Eq.~\ref{eq:jdot} is the velocity difference at the NS surface.

We adopt the accretion model from \citet{Debatri2020} which is fitted to Figure 4 of \citet{MacLeod:2015ApJ}, with $\Delta M_{\rm NS} \le 0.1 M_{\odot}$, e.g.
\begin{equation}
    \Delta M_{\rm NS}/M_{\odot} = a(R_{\rm comp}/R_{\odot}) + b.
\end{equation}

\section{N-body Simulation and Preliminary Results}

We implemented pulsar evolution in \textsc{NBODY6++GPU} by introducing the switch, \texttt{KZ(29)}, which enables or disables the pulsar treatment.  
The code accepts pulsar parameters through the namelist input file, including the birth distributions of spin period and magnetic field, as well as the magnetic field decay timescale and mass scale.

As a first test, we model the Galactic globular cluster M71.  
M71 has a present-day mass of $1.7 \times 10^4\,M_{\odot}$, an age of $\sim 10$\,Gyr, and hosts five observed pulsars, all in binary systems.  
To mimic its properties, we initialised a cluster with 200,000 particles, adopting a $10\%$ primordial binary fraction and metallicity $Z=0.002$.  
Stellar masses were drawn from a Kroupa initial mass function \citep{Kroupa2001} ranging $0.08$--$150\,M_{\odot}$.  
Initial conditions were generated using \textsc{mcluster} \citep{Kamlah2022}.  
We also use a NS natal kick distribution with Maxwellian distribution peaking at 10 km/h for retaining NSs in the cluster. 

\begin{figure}
    \centering
    \includegraphics[width=0.45\linewidth]{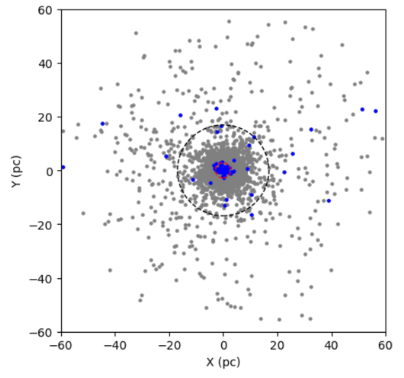}
    \includegraphics[width=0.45\linewidth]{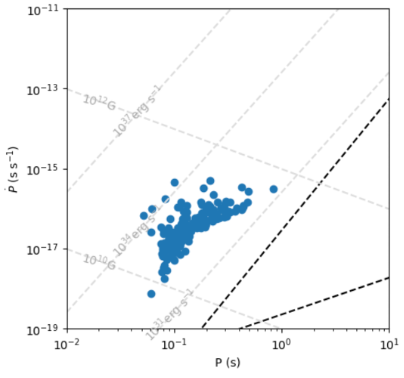}
    \caption{
    Left: Spatial distribution of cluster members (gray) and pulsars (blue). The black dashed circle marks the half-mass radius, and the red dashed circle the core radius.  
    Right: $P$–$\dot{P}$ diagram of pulsars at 20 Myr. The population includes 179 isolated pulsars and 2 pulsars in binaries.}
    \label{fig:fig}
\end{figure}

At the time of the Symposium, the cluster has been evolved for 20 Myr with the simulation.  
At this snapshot, the simulation contains 179 isolated pulsars and 2 pulsars in binaries.  
Figure~\ref{fig:fig} shows both the spatial distribution of cluster members and the $P$–$\dot{P}$ diagram of the pulsars.

\section{Follow-up}

Once the \textit{N}-body simulation of M71 progresses to $\sim 10$\,Gyr, we will examine the resulting pulsar population in detail.
We will combine the intrinsic pulsar population with survey parameters for current and future radio facilities (e.g., FAST, MeerKAT, SKA) to predict the number and properties of observable pulsars.  
We will also study double compact object mergers as gravitational wave sources and compare them with detections from the LVK collaborations.
Finally, we will use the simulated pulsars to explain observed $\gamma$-ray emission from M71.


\begin{thebibliography}{}


\bibitem[Chattopadhyay et al.(2020)]{Debatri2020} Chattopadhyay, D., Stevenson, S., Hurley, J.~R., et al.\ 2020, Monthly Notices of the Royal Astronomical Society, 494, 2, 1587. doi:10.1093/mnras/staa756

\bibitem[Dirson et al.(2022)]{Dirson2022} Dirson, L., P{\'e}tri, J., \& Mitra, D.\ 2022, Astronomy \& Astrophysics, 667, A82. doi:10.1051/0004-6361/202243305

\bibitem[\protect\citeauthoryear{Eggleton}{1983}]{1983ApJ...268..368E} Eggleton P.~P., 1983, the Astrophysical Journal, 268, 368. doi:10.1086/160960

\bibitem[Henry et al.(2024)]{Henry2024} Henry, O.~K., Paglione, T.~A.~D., Song, Y., et al.\ 2024, Monthly Notices of the Royal Astronomical Society, 535, 1, 434. doi:10.1093/mnras/stae2402

\bibitem[Johnston et al.(2020)]{Johnston2020} Johnston, S., Smith, D.~A., Karastergiou, A., et al.\ 2020, Monthly Notices of the Royal Astronomical Society, 497, 2, 1957. doi:10.1093/mnras/staa2110

\bibitem[Kamlah et al.(2022)]{Kamlah2022} Kamlah, A.~W.~H., Leveque, A., Spurzem, R., et al.\ 2022, Monthly Notices of the Royal Astronomical Society, 511, 3, 4060. doi:10.1093/mnras/stab3748

\bibitem[Kiel et al.(2008)]{Kiel:2008MNRAS} Kiel, P.~D., Hurley, J.~R., Bailes, M., et al.\ 2008, Monthly Notices of the Royal Astronomical Society, 388, 1, 393. doi:10.1111/j.1365-2966.2008.13402.x

\bibitem[Konar \& Bhattacharya(1997)]{Konar:1997MNRAS.284..311K} Konar, S. \& Bhattacharya, D.\ 1997, Monthly Notices of the Royal Astronomical Society, 284, 2, 311. doi:10.1093/mnras/284.2.311


\bibitem[Kroupa(2001)]{Kroupa2001} Kroupa, P.\ 2001, Monthly Notices of the Royal Astronomical Society, 322, 2, 231. doi:10.1046/j.1365-8711.2001.04022.x

\bibitem[Lattimer \& Schutz(2005)]{lattimer2005} Lattimer, J.~M. \& Schutz, B.~F.\ 2005, the Astrophysical Journal, 629, 2, 979. doi:10.1086/431543


\bibitem[MacLeod \& Ramirez-Ruiz(2015)]{MacLeod:2015ApJ} MacLeod, M. \& Ramirez-Ruiz, E.\ 2015, the Astrophysical Journal, 798, 1, L19. doi:10.1088/2041-8205/798/1/L19

\bibitem[Romani(1990)]{Romani:1990Natur} Romani, R.~W.\ 1990, Nature, 347, 6295, 741. doi:10.1038/347741a0


\bibitem[Song et al.(2024)]{Song2024} Song, Y., Stevenson, S., Chattopadhyay, D, et al.\ 2024, , arXiv:2406.11428. doi:10.48550/arXiv.2406.11428


\bibitem[Spurzem \& Kamlah(2023)]{Spurzem2023} Spurzem, R. \& Kamlah, A.\ 2023, Living Reviews in Computational Astrophysics, 9, 1, 3. doi:10.1007/s41115-023-00018-w

\bibitem[Tauris \& van den Heuvel(2006)]{Tauris2006} Tauris, T.~M. \& van den Heuvel, E.~P.~J.\ 2006, Compact stellar X-ray sources, 39, 623. doi:10.48550/arXiv.astro-ph/0303456


\bibitem[Wang et al.(2015)]{Wang2015} Wang, L., Spurzem, R., Aarseth, S., et al.\ 2015, Monthly Notices of the Royal Astronomical Society, 450, 4, 4070. doi:10.1093/mnras/stv817

\bibitem[Ye et al.(2019)]{Ye2019ApJ} Ye, C.~S., Kremer, K., Chatterjee, S., et al.\ 2019, the Astrophysical Journal, 877, 2, 122. doi:10.3847/1538-4357/ab1b21

\bibitem[Young et al.(2010)]{Young:2010MNRAS} Young, M.~D.~T., Chan, L.~S., Burman, R.~R., et al.\ 2010, Monthly Notices of the Royal Astronomical Society, 402, 2, 1317. doi:10.1111/j.1365-2966.2009.15972.x

\bibitem[Zhang \& Kojima(2006)]{Zhang:2006MNRAS} Zhang, C.~M. \& Kojima, Y.\ 2006, Monthly Notices of the Royal Astronomical Society, 366, 1, 137. doi:10.1111/j.1365-2966.2005.09802.x


\end{thebibliography}
\end{document}